\documentclass[12pt]{article}

\begin{document}

\begin{center}
{\LARGE Consequences of 't Hooft's Equivalence Class Theory and Symmetry by }%
{\Large Coarse Graining}

X.F.Liu $^{a,b}$ and C.P.Sun$^b,$\footnote{%
e-amil:suncp@itp.a.c.cn; web:http://www.itp.ac.cn/~suncp}

$^a$Department of Mathematics, Peking University, Beijing, 100871,China

\bigskip $^b$Institute of Theoretical Physics, Chinese Academy of Sciences,
Beijing 100080, China

\textbf{Abstract}
\end{center}

\begin{quotation}
\textit{According to 't Hooft (Class.Quantum.Grav. 16 (1999), 3263), quantum
gravity can be postulated as a dissipative deterministic system, where
quantum states at the ``atomic scale''can be understood as equivalence
classes of primordial states governed by a dissipative deterministic
dynamics law at the ``Planck scale''. In this paper, it is shown that for a
quantum system to have an underlying deterministic dissipative dynamics, the
time variable should be discrete if the continuity of its temporal evolution
is required. Besides, the underlying deterministic theory also imposes
restrictions on the energy spectrum of the quantum system. } \textit{It is
also found that quantum symmetry at the ``atomic scale'' can be induced from
't Hooft's Coarse Graining classification of primordial states at the}
``Planck scale".
\end{quotation}


\section{Introduction}

Recently , Gerard't Hooft postulated that there should be a dissipative
deterministic theory underlying quantum gravity at the so called ``Planck
scale''[1,2]. In his theory, the generic quantum mechanics is no longer the
crucial starting point. Rather, a deterministic theory with dissipation of
information at the ``Planck scale'' is needed to derive quantum mechanics at
the ``atomic scale''.It seems that this viewpoint can solve problems
concerning locality and causality in the so called Planck scale physics such
as quantum gravity, which are quite different from those in the usual
quantum field theories in some flat background space-time based on the
holographic principle in quantum gravity theory [3].

In 't Hooft's opinion, at the ``atomic scale'' quantum states are
equivalence classes of primordial states at the ``Planck scale''. In
Ref.[4],this point of view was illustrated through a simple model. According
to 't Hooft, if we only care the temporal evolution of equivalence classes,
the information within each equivalence class could be ignored. Then from a
non-time-reversible evolution, which characterizes a deterministic process
with dissipation at the ``Planck scale'', we can obtain a time-reversible
evolution of the properly defined equivalence classes of primordial states.
Taking the equivalence classes to be quantum states we are then able to
introduce a reversible evolution law at the ``atomic scale''. Apparently,
here the central problem is how to classify the Planck scale states with
respect to a deterministic evolution.'t Hooft's solution to this problem is
as follows. He argues that two Planck scale states are equivalent at the
``atomic scale'' if , after some finite time interval, they evolve into the
same state. This leads to a natural definition of equivalence classes: two
states are in the same equivalence class if and only if they evolve into the
same state after some finite time interval. Then, quantum states are
identified with these equivalence classes.

Most recently we make clear the mathematical structure of 't Hooft's theory
using quotient space construction and the related concepts [5]. Let the
primordial states span a linear space . We find that the equivalence classes
defined by 't Hooft can be identified with the cosets of the invariant
subspace spanned by those primordial states annihilated by the
time-evolution operator. Thus the Hilbert space of quantum states is just
the corresponding quotient space and the time-reversible evolution at the
``atomic scale'' can naturally be induced on the quotient space by the
dissipative deterministic evolution operator . Following this line, in this
paper, we will make a further analysis of the mathematical aspect of 't
Hooft's theory and then discuss some physical consequences implied in the
theory. We will also probe the spectral structure of finite dimensional
quantum system with an underlying deterministic structure and extend 't
Hooft's idea to study quantum symmetry problem.

\section{Some Mathematical Results}

In this section we present some mathematical results closely related to the
't Hooft equivalence class theory. In the following, $I,J$ stands for index
sets not necessarily finite;if $V_1$ is a subspace of $V,v\in V,$the element
$v+V_1$ in the quotient space $V/V_1$ is denoted by $\overline{v}.$ All the
vector spaces to be considered are over the complex number field. Physically
, one should bear in mind that $V$ will be the linear space spanned by
so-called primordial states at the ``Planck scale''(see below).For
convenience,we list the mathematical definitions of some concepts appearing
in 't Hooft's theory as follows.

\textbf{Definition 1.} A linear operator $T\in End(V)$ is called
deterministic if there exists a basis $\{v_i|i\in I\}$ of $V$ on which $T$
acts in the following way:$\forall i\in I,\exists i^{\prime }\in I\ \ s.t.\
Tv_i=v_{i^{\prime }}.$ Such a basis is called $T-$deterministic basis. If,
moreover, $T$ is singular (non-invertible), then it is called dissipative
deterministic.

\textbf{Remark 1.} In 't Hooft theory, $T$ represents a deterministic
time-evolution process (with dissipation) at the ``Planck scale''.

\textbf{Definition 2.} An injective map from a set to itself is called a
permutation of the set. A linear operator $T\in End(V)$ is called a
permutation operator if there exists a basis of $V$ on which $T$ act as a
permutation.

\textbf{Definition 3.} A linear operator on a vector space is called
unitarizable if there exist an inner product on the vector space such that
it is unitary relative to it.

\textbf{Remark 2.} Physically, time-reversible evolution is described by an
unitary operator,and a reversible but not unitarizable operator usually does
not correspond to any practical evolution in quantum mechanics.

\textbf{Definition 4.} Let $V$ and $W$ be two vector spaces, $T\in End(V)$
and $S\in End(W)$. If there exists an isomorphism $\varphi $ between $V$ and
$W$ such that $\varphi T=S\varphi $, $T$ and $S$ are called equivalent.

Having prepared the above definitions, we now state one of our central
results .

\textbf{Proposition 1.} Let $V$ be a vector space, $T\in End(V)$ is
dissipative deterministic and $V_1$ is a $T-$invariant subspace such that
the induced operator $\overline{T} $ on the quotient space $V/V_1$ is
nonsingular, then $\overline{T}$ is a permutation operator; Conversely, if $%
S $ $\in End(V)$ is a permutation operator, then there exists a vector space
$V^{\prime },$a dissipative deterministic operator $S^{\prime }\in
End(V^{\prime })$ and a $S^{\prime }-$invariant subspace $V_1^{\prime }$ of $%
V^{\prime }$ such that the induced operator $\overline{S^{\prime }}\in
End(V^{\prime }/V_1^{\prime })$ is equivalent to $S.$

Proof. Let $\{v_i|i\in I\}$ be a $T-$deterministic basis. Then there exists
a subset $J\subset I$ such that $\{\overline{v}_i|i\in J\}$ is a basis of $%
V/V_1.$By definition
\begin{equation}
\overline{T}\overline{v}_i=\overline{Tv_i}=\overline{v}_{i^{\prime
}}\;(i,i^{\prime }\in I).
\end{equation}
As $\overline{T}$ is nonsingular,we clearly see that $\overline{T}$ acts as
a permutation on the basis $\{\overline{v}_i|i\in J\}.$This proves the first
half of the proposition. For the second half,let $\{v_i|i\in I\}$ be a basis
of $V$ on which $S$ acts as a permutation, take an arbitrary element $%
w\notin V$ and define
\[
V^{\prime }=span\{v_i,w|i\in I\}
\]
$,V_1^{\prime }=V.$Define $S^{\prime }\in End(V^{\prime })$ such that $%
S^{\prime }|_V=S$ and $S^{\prime }w=0.$It is then trivial to verify that $%
S^{\prime }$ is dissipative deterministic and $\overline{S^{\prime }}$ is
equivalent to $S$.The proposition is thus proved.

\textbf{Remark 3.} This proposition,as we will see below, tells us that 't
Hooft 's underlying dissipative deterministic dynamic law at the ``Planck
scale'' can only produce very special time-reversible evolution at the ''
atomic scale'' .

Keep the notations in the above proposition. We have the following corollary.

\textbf{Corollary.} If $V/V_1$ is finite dimensional, then $\overline{T}$ is
unitarizable.

Proof. According to the proposition, there is a basis of $V/V_1$ on which $%
\overline{T}$ acts as a permutation. If $\dim V/V_1<\infty ,\overline{T}$ is
periodic,namely,there exists a positive integer $n$ such that $\overline{T}%
^n=1.$Let $p$ be its period. Choose an arbitrary inner product $(\ ,\ )$ on $%
V/V_1$ and define a new inner product $<\,,\,>$ as follows:
\begin{equation}
\left\langle \overline{v},\overline{w}\right\rangle =\sum_{j=1}^p(\overline{T%
}^j\overline{v},\overline{T}^j\overline{w}),\forall \,\overline{v},\overline{%
w}\in V/V_1.
\end{equation}
It is then easy to show that $\overline{T}$ is unitary relative to the inner
product $<\,,\,>.$

Proposition 1 shows us that an invertible linear operator can be induced
from a deterministic operator $T$ if and only if it is a permutation
operator. The following proposition characterizes permutation operator on a
finite dimensional space.

\textbf{Proposition 2.} Let $V$ be a finite dimensional vector space, $T\in
End(V).$ $T$ is a permutation operator if and only if it is diagonalizable
and its eigenvalues can be grouped into some classes, say, $\Delta
_{n_1},\Delta _{n_2},\cdots ,\Delta _{n_r},$such that $\Delta
_{n_j}(j=1,2,\cdots ,r)$ exactly consists of the $n_j$ $n_j$th roots of
unity with the same multiplicity.

Proof. Let $\{v_i|i=1,2,\cdots ,n\}$ be a basis on which $T$ acts as a
permutation. First,suppose $T$ is a cyclic permutation the basis,namely,we
have
\begin{equation}
Tv_1=v_2,Tv_2=v_3,\cdots ,Tv_{n-1}=v_n,Tv_n=v_1.
\end{equation}
Then $T$ is a periodic operator of period $n,$and its minimal polynomial is $%
\lambda ^n-1.$Consequently, $T$ is diagonalizable and its eigenvalues are $%
e^{i\frac{2k\pi }n}$ $(k=1,2,\cdots ,n),$the $nth$ roots of unity. Now let $%
T $ act as a general permutation on the basis. We notice that the basis
elements can be grouped into some classes on each of which $T$ acts as a
cyclic permutation. Thus the ``only if'' part of the proposition easily
follows.

Conversely,suppose $T$ is diagonalizable and its eigenvalues can be grouped
into some classes $\Delta _{n_1},\Delta _{n_2},\cdots ,\Delta _{n_r}$ in
such a way that $\Delta _{n_j}(j=1,2,\cdots ,r)$ exactly consists of the $%
n_j $ $n_jth$ roots of unity with multiplicity $m_j.$Then there is a basis $%
\{v_{k,l}^j|j=1,2,\cdots ,r;k=1,2,\cdots ,n_j;l=1,2,\cdots ,m_j\}$ such that
\begin{equation}
Tv_{k,l}^j=e^{i\frac{2k\pi }{n_j}}v_{k,l}^j.
\end{equation}
Define the subspace $V_{j,l}$ of $V$ as follows:
\[
V_{j,l}=span\{v_{k,l}^j|k=1,2,\cdots ,n_j\}.
\]
Clearly, we have
\[
V=\sum_{j=1}^r\sum_{l=1}^{m_j}\oplus V_{j,l}
\]
and from the proof of the ``only if'' part we know in each subspace $V_{j,l}$
there is a basis on which $T$ acts as a cyclic permutation of order $m_j$.
Put together,these bases of the subspaces form a basis of $V$ on which $T$
acts as a permutation. This proves the ``if'' part of the proposition.

\section{Dynamics from 't Hooft's Theory}

In this section we focus on the physical aspect of 't Hooft's theory, but
our analysis depends on the above mathematical results.

In 't Hooft's theory, primordial states at the ``Planck scale'' need not
form a linear space. Generally they can be denoted by a set $\Sigma $ $%
=\{\phi _i|i\in I\}$. \textit{The underlying deterministic evolution is a
transformation }$U$\textit{\ (usually depending on time)of }$\Sigma $\textit{%
\ to itself. If }$U$\textit{\ has no inverse it is called a dissipative
deterministic evolution. }Obviously\textit{, }it can be represented by a
matrix with the entries $0$ or $1$ if $I$ is a finite set. As $U$ is an
evolution operator ,we write it as $U=U(t_f,t_{i,})$ by convention.
Physically, it represents the evolution in the time interval $[t_i,t_f].$
Certainly the evolution should satisfy the so called semi-group condition
\begin{equation}
U(t_f,t_m)U(t_m,t_i)=U(t_f,t_i)
\end{equation}
\[
U(t,t)=1
\]
If $U$ is singular, it describes deterministic process with dissipation. As
a matter of fact, under such an evolution some states will disappear and
some states will evolve into the same state,or in other words, some states
with a different past may have the same deterministic fate.'t Hooft thinks
that if two states evolve in such a way that their futures are identical
they should represent the same state at the ``atomic scale''.In this view,he
divides the elements of $\Sigma $ into equivalence classes, $\phi _{i_1}$
and $\phi _{i_2}$ ($i_{1,}i_2\in I$) being in the same equivalence class if
they are evolved into the same state after finite time interval. Denote by $%
\Xi =\{\overline{\phi }_j|j\in J\}$ the set of the equivalence classes. Then
't Hooft postulates that the space of quantum states is spanned by $\{%
\overline{\phi }_j|j\in J\}$ and claims that the reduced evolution on the
space of quantum states is reversible. We can mathematically reformulate 't
Hooft's theory as follows[5].We assume that the evolution operator $%
U(t_2,t_1)$ only depend on the difference of $t_2$and $t_1$, i.e., we can
write $U(t_2,t_1)=U(t_2-t_1)$ .This is in the spirit of 't Hooft's original
construction. Then the evolution at the ``Planck scale'' is determined by
the operator $U(t,0)\stackrel{\triangle}{=} U(t).$ Let $V$ be the vector
space spanned by $\{\phi _i|i\in I\}.$Then $U(t)$ can be extended to \textit{%
a deterministic operator} on $V$. We call $V$ the space of primordial states
in spite of the fact that generally it contains elements which are not
states. Let $V_1$ denote the subspace of $V$ consisting of the vectors
annihilated by $U(t)$ at some $t$, namely, a vector $v$ belongs to $V_1$if
and only if there exists some $U(t)$ such that $U(t)v=0$. Then it follows
that \textit{the space of quantum states is none other than the quotient
space }
\[
Q=V/V_1=\{|\phi \rangle \stackrel{\triangle}{=} \phi +V_1|\phi \in V\}
\]
and a non-singular evolution law of the quantum states naturally follows
from $U(t).$ Let $\overline{v}$ $\equiv |\nu \rangle $ denote the
equivalence class containing $v$ . We notice that $V_1$is invariant under $%
U(t)$. Thus $U(t)$ induces a natural action on the quotient space $Q.$We
denote the induced operator by $\overline{U(t)},$ then we have
\begin{equation}
\overline{U(t)}\overline{v}=\overline{U(t)v}.
\end{equation}
The following simple result is easy to prove.

\textbf{Proposition 3.} $\overline{U(t)}$ is non-singular.

In fact, if $\overline{U(t)}\overline{v}=\overline{0}$ , then $U(t)v\in V_1.$%
Thus there exists some $t^{\prime }$ such that $U(t^{\prime })U(t)v=0.$It
then follows that
\begin{equation}
U(t^{\prime })U(t)v=U(t^{\prime }+t)v=0
\end{equation}
By definition this means $v\in V_1,$i.e.,$\overline{v}=\overline{0}$.This
proves the non-singularity of $\overline{U(t)}.$

\textbf{Remark 4.} In Ref[1,2], 't Hooft just claims the non-singularity of
the induced evolution operator. But it should be pointed out that if the
condition $U(t_2,t_1)=U(t_2-t_1)$ is not satisfied the induced evolution $%
\overline{U(t_2,t_1)}$ might be singular if we still use 't Hooft's
principle to classify the primordial states.

We are now in a position to discuss a consequence of 't Hooft's theory. The
basis consisting of the equivalence classes is called the primordial basis
by 't Hooft. In our notations,$\{\overline{\phi }_j|j\in J\}$ is the
primordial basis and $U(t)$ is a (dissipative) deterministic operator on $V.$
As we have proved the non-singularity of $\overline{U(t)},$it follows from
Proposition 1 that $\overline{U(t)}$ is a permutation operator which acts as
a permutation on the primordial basis. Then we easily observe that if we
require $\overline{U(t)}$ to be continuous with respect to $t$, the time
variable should be discrete. For example,if $J$ is a finite set,or in other
words,the quantum Hilbert space is finite dimensional,the induced evolution
operator $\overline{U(t)}$ is represented as a matrix with the entries $0$
or $1$ with respect to the primordial basis. Clearly,it could not be
continuous if the time variable is not discrete.

\section{Spectrum and Hamiltonian}

Let us turn to consider restrictions on the energy spectrum of quantum
system imposed by the underlying determinism. Due to the arguments in the
last paragraph,we assume the time variable to be discrete. Without losing
generality, let the time $t$ take values in $Z^{+},$the set of non-negative
integers. The deterministic evolution and the induced evolution of the
quantum system is then completely determined by the operator $U(1).$ Suppose
$\overline{U(1)}$ is unitary. It is then can be written as $\overline{U(1)}%
=e^{-iH},$where $H$ is a Hermitian operator describing the Hamiltonian of
the quantum system. Now if the quantum system is finite dimensional it
follows from Proposition 2 that the eigenvalues of $\overline{U(1)}$ are of
the form $e^{-i\frac{2k\pi }n}.$Thus we have the following

\textbf{Proposition 4.} The eigenvalues of $H$ corresponding to the induced
evolution $\overline{U(1)}=e^{-iH}$ of quantum states lie in the set
\[
\{\frac{2k\pi }n\pm 2m\pi |k,n,m\in Z^{+}\}.
\]

\textbf{Remark 5.} We have seen that evolutions that can be induced from
dissipative deterministic evolutions at the ``Planck scale'' belong to a
special class. Firstly, there is a rather strict restriction on the
corresponding Hamiltonian $H$ . Secondly, if a quantum system with an
underlying deterministic structure as is described by 't Hooft is initially
in the state represented by an element of the primordial basis then the
evolution will never cause coherent superposition of quantum states. As
these drawbacks are inherent in the theory, to remove them we have to
generalize the underlying dynamic law at the ``Planck scale''.

Another conclusion that can be drawn from Proposition 1 is that 't Hooft's
theory is closely related to the hidden variable theory. Since $\overline{%
U(t)}$ acts as a permutation on the primordial basis of the space of quantum
states, an operator that is diagonal now with respect to this basis will
continue to be diagonal in the future. Such an operator could thus be
thought to represent a hidden variable. This suggests that a quantum system
with an underlying dissipative deterministic mechanism might permit some
kind of hidden variable theory. The corollary to Proposition 2 also shows us
that if $U(t)$ is a dissipative deterministic such that the quotient space $%
V/V_1$ is finite dimensional $\overline{U(t)}$ can be made unitary by
properly introducing an inner product. Then $\overline{U(t)}$ can be
regarded as an evolution operator for a quantum system. But on the other
hand, such inner product is not at all unique. Since a correct quantum
theory requires a Hilbert space with properly defined inner product to
define probability, this is really a problem if we wish to derive quantum
dynamics from a dissipative deterministic evolution, not just to interpret a
given quantum system as governed by an underlying deterministic mechanism.
So a gap remains to be bridged between the so called Planck scale physics
and the atomic scale physics in 't Hooft's theory.

Before passing to discuss quantum symmetry we would like to present a simple
quantum system which has some characteristics of a deterministic system as
shown above. We consider the following quantum system: A spinless free
particle in the one dimensional region $[0,L]$ with the boundary condition $%
\psi (0,t)=\psi (L,t)$,where $\psi (x,t)$ is the wave function. The Hilbert
space of the system is
\[
\mathcal{H=}\{\psi \in \mathcal{L}^2[0,L]|\psi (0)=\psi (L)\}.
\]
Clearly,
\[
\triangle =\{e^{i\frac{2k\pi }Lx}|k=0,\pm 1,\pm 2,\cdots \}
\]
is a basis of $\mathcal{H}.$ In the case of extreme relativity, the
Hamiltonian of the system is $H=-i\hbar c\frac d{dx},$where $c$ is the speed
of light. Define $\overline{U(t)}=e^{-iHt}.$We have
\begin{equation}
\overline{U(t)}e^{i\frac{2k\pi }Lx}=e^{-i2k\pi \frac{\hbar c}Lt}e^{i\frac{%
2k\pi }Lx}
\end{equation}
We observe that if we take the time to be discrete, it is then possible to
define a time unit such that the one step evolution acts on $\triangle $ in
the following way:
\[
\overline{U(1)}e^{i\frac{2k\pi }Lx}=e^{i\frac{2k\pi }Lx}
\]
We then see that this system might be regarded as a deterministic system and
$\triangle $ might serve as primordial basis for the system. If we normalize
$\frac{\hbar c}L$ as one energy unit, then the energy spectrum of the system
is$\{2k\pi |k=0,\pm 1,\pm 2,\cdots \}.$This is consistent with our
discussion above.

\textbf{Remark 6.} It should be pointed out that the above simple example is
essentially the same as the example of massless neutrinos discussed in
Ref.[1].

\section{Quantum Symmetry by Coarse Graining}

As shown above, 't Hooft's classification of primordial states implies a
scheme for coarse graining. Usually, for a large close system a coarse
graining process can result in quantum dissipation and decoherence in the
subsystem [6]. But here the converse seems to be the case: coarse graining
(or classification) can lead to a unitary dynamics for the effective system
even if the evolution of primordial system is not time-reversible. Since
``symmetry dominates dynamics'', it is rather natural to probe the role of
coarse graining in generating symmetry at the ``atomic scale''.

Let a deterministic system be described by an evolution operator $\overline{%
U(t)}$, and let $\{\overline{\phi }_j|j\in J\}$ be the primordial basis for
the system. Denote by $P^J$ the permutation group of the set $J.$ According
to Proposition 1, $\overline{U(t)}$ is a permutation operator and can be
identified with an element of $P^J.$ By definition, the group of quantum
symmetry consists of those unitary operators on the state space that commute
with the evolution operator. If we require that these unitary operators be
induced from deterministic operators on the space of primordial states, it
then follows from Proposition 1 that they belong to the centralizer of $%
\overline{U(t)}$ in $P^J.$ If the space of quantum states is finite
dimensional, by the trick of redefining inner product as is used in the
proof of the corollary to Proposition 1,we can show that there exists an
inner product such that both $\overline{U(t)}$ and the operators in its
centralizer in $P^J$ are unitary operators. Thus in this case it might be
reasonable to take the group of quantum symmetry to be the centralizer of $%
\overline{U(t)}$ in $P^J.$Anyway, the symmetry group is a discrete group.

We have seen that if we adhere to the principle that things happening in the
space of primordial states bear the mark of determinism, then
logically,things happening in the space of quantum states bear the mark of
discreteness. To change the situation we need to loosen the restriction of
determinism in the strict sense of this word used by 't Hooft. Let us
conclude this paper by a short discussion of quantum symmetry derived from a
not necessarily deterministic operator on the space of primordial states.
Let $V$ be the space of primordial states and $S\in End(V)$ satisfies $%
SV_1\subset V_1.$

\textbf{Proposition 5.} $\overline{U(t)}\overline{S}-\overline{S}\overline{%
U(t)}=0$ if and only if there exists some $t^{\prime }$ such that
\begin{equation}
U(t^{\prime })(U(t)S-SU(t))=0
\end{equation}

The proof of this result is immediate. It directly follows from eq.(9) that $%
(U(t)S-SU(t))V\subset V_1$(cf. Section 3).In other words,we have
\begin{equation}
\overline{U(t)}\overline{S}-\overline{S}\overline{U(t)}=0
\end{equation}
This proves the ``if'' part. The ``only if'' part can be proved by reversing
the deduction.

If the time is discrete and takes values in $Z^{+},$then the evolution at
the ``Planck scale'' is determined by $U(1)\stackrel{\triangle}{=} U.$Notice
that $U^n=U(n)$ in this case.It follows that eq.(9) is equivalent to the
following equation
\begin{equation}
U^n(US-SU)=0
\end{equation}
for some positive integer $n.$Let us take 't Hooft's example in Ref.[1] to
illustrate the above idea. We have
\begin{equation}
U=\left(
\begin{array}{cccc}
0 & 1 & 0 & 1 \\
1 & 0 & 0 & 0 \\
0 & 0 & 1 & 0 \\
0 & 0 & 0 & 0
\end{array}
\right) .
\end{equation}
Let $e_1=\left(
\begin{array}{cccc}
1 & 0 & 0 & 0
\end{array}
\right) ^T,e_2=\left(
\begin{array}{cccc}
0 & 1 & 0 & 0
\end{array}
\right) ^T,e_3=\left(
\begin{array}{cccc}
0 & 0 & 1 & 0
\end{array}
\right) ^T,$

$e_4=\left(
\begin{array}{cccc}
0 & 0 & 0 & 1
\end{array}
\right) ^T.$Then
\[
\overline{U}=\left(
\begin{array}{ccc}
0 & 1 & 0 \\
1 & 0 & 0 \\
0 & 0 & 1
\end{array}
\right)
\]
with respect to the basis $\{$ $\overline{e_i}|i=1,2,3\}.$The general matrix
$T$ that commutes with $\overline{U}$ is of the form
\[
T=\left(
\begin{array}{ccc}
a & b & c \\
b & a & c \\
j & j & k
\end{array}
\right) .
\]
Suppose $SV_1\subset V_1.$ Then the general solution of eq.(11) is
\[
S=\left(
\begin{array}{cccc}
a & b & c & b \\
b-m & f & g & f \\
j & j & k & j \\
m & a-f & c-g & a-f
\end{array}
\right)
\]
It is clear that for each $T$ commuting with $\overline{U}$ there exists $S$
such that $\overline{S}=T.$In fact, as $\overline{e_2}=\overline{e_4}$, the
above $S$ has the representation
\begin{equation}
\overline{S}=\left(
\begin{array}{ccc}
a & b & c \\
b & a & c \\
j & j & k
\end{array}
\right)
\end{equation}
with respect to the basis $\{$ $\overline{e_i}|i=1,2,3\}.$
Mathematically,this is a trivial fact. On the other hand,we have
\[
US-SU=\left(
\begin{array}{cccc}
0 & 0 & 0 & 0 \\
a-f & m & c-g & m \\
0 & 0 & 0 & 0 \\
-a+f & -m & -c+g & -m
\end{array}
\right) .
\]
This simply means that for $\overline{S}$ to commute with $\overline{U}$ , $%
S $ does not necessarily commute with $U.$

In the representation where $\overline{U}$ is diagonal we have
\[
\overline{U}=\left(
\begin{array}{ccc}
1 & 0 & 0 \\
0 & 1 & 0 \\
0 & 0 & -1
\end{array}
\right).
\]
Thus the matrix $T$ that commutes with $\overline{U}$ takes a block diagonal
form.It then follows that the symmetry group of the system is $U(2)\times
U(1)$.But if we impose the restriction that $S$ is a deterministic
operator,as is required by determinism,it then turns out that the set of
nonsingular $\overline{S}$ commuting with $\overline{U}$ is $\{1,\overline{U}%
\},$the centralizer of $\overline{U}$ in $P^3.$

To sum up,if we loosen the restriction of determinism it is possible to
induce quantum symmetry from transformations on the space of primordial
states through a procedure of coarse graining as shown above. On the other
hand,quantum symmetry at the ``atomic scale'' does not necessitate symmetry
at the ``Planck scale''.

\medskip

\noindent \textit{\ This work is supported by the NFS of China. The authors
would like to express their thanks to S. X. Yu, Y. L. Wu, M.Yu and M.Li for
helpful discussion with them.}

\end{document}